# Review and Examination of Input Feature Preparation Methods and Machine Learning Models for Turbulence Modeling

Shirui Luo [a], Jiahuan Cui [b], Madhu Vellakal [a], Jian Liu [b], Enyi Jiang [b], Seid Koric [a], Volodymyr Kindratenko [a]

[a]) *National Center for Supercomputing Applications, University of Illinois at Urbana-Champaign, Urbana, IL 61801, USA,*

[b]) *Zhejiang University – University of Illinois at Urbana-Champaign Institute, Haining, Zhejiang province 314400, China*

## Abstract

Model extrapolation to unseen flow is one of the biggest challenges facing data-driven turbulence modeling, especially for models with high dimensional inputs that involve many flow features. In this study we review previous efforts on data-driven Reynolds-Averaged Naiver Stokes (RANS) turbulence modeling and model extrapolation, with main focus on the popular methods being used in the field of transfer learning. Several potential metrics to measure the dissimilarity between training flows and testing flows are examined. Different Machine Learning (ML) models are compared to understand how the capacity or complexity of the model affects its behavior in the face of dataset shift. Data preprocessing schemes which are robust to covariate shift, like normalization, transformation, and importance re-weighted likelihood, are studied to understand whether it is possible to find projections of the data that attenuate the differences in the training and test distributions while preserving predictability. Three metrics are proposed to assess the dissimilarity between training/testing dataset. To attenuate the dissimilarity, a distribution matching framework is used to align the statistics of the distributions. These modifications also allow the regression tasks to have better accuracy in forecasting under-represented extreme values of the target variable. These findings are useful for future ML based turbulence models to evaluate their model predictability and provide guidance to systematically generate diversified high-fidelity simulation database.

## 1. Introduction

Accurate and computationally feasible turbulent flows simulation is of critical importance to many practical applications, such as turbomachinery, mixing and combustion of reacting gases, atmospheric re-entry vehicles, and commercial aircraft design. Despite the growth in computational power over the last decade, modeling and simulation of turbulent flows to a desired level of accuracy is still challenging. High fidelity simulations such as Direct Numerical Simulations (DNS) and Large Eddy Simulation (LES) continue to be largely computationally inaccessible for industrial enterprises.[1] Reynolds-averaged Navier–Stokes (RANS) simulation keeps being a workhorse Computational Fluid Dynamics (CFD) method due to its computational efficiency and easy implementation.[2] Although scale-resolved predictions give better results, continued efforts to expand the affordable range of applicability and accuracy of RANS models remains crucial to the further evolution and application of CFD for industry.[2] Turbulence modeling for the unclosed terms in RANS equations has traditionally evolved through a combined efforts of mathematics, flow theory, empiricism, and rudimentary data-driven techniques such as single or two variable curve-fitting. However, RANS only resolves the largest energy containing motions of the flow, which results in unsatisfactory predictive accuracy in many complex flows or flow in different contexts.[3] Many improvements have been proposed, but their use in the industry is still limited due to the need to tune many undetermined parameters based on dataset from particular classes of problems.[4-6] Industry needs affordable predictions with defined uncertainty in an ever expanding range of conditions and geometric complexity.

Recent ML-based methods, which can comprehensively utilize high-fidelity simulation data for improving turbulence modeling in RANS simulations, have gained significant interest due to ML's remarkable approximation power and the increased availability of large data sets.[7-9] The main concepts behind ML-based turbulence modeling is that instead of looking for the complete closure form, we could presumably postulate a relationship between the unclosed terms with other known features and use neural network to fit unknown parameters. A comprehensive review of data-driven approaches that contain statistical inference and ML for turbulence modeling is presented in a recent article.[10] The integration of ML with turbulence models is generally accomplished via three distinct strategies: (1) modeling the Reynolds stress anisotropic tensor directly from input features either derived from physics intuition[11,12] or Galilean invariants (*e.g.* mean strain rate and rotation rate tensor)[13-15]; (2) modeling the deficiencies in the functional form of the production, destruction, and diffusion terms in turbulence transport equations[16-18]; (3) modeling new correction terms as the discrepancy between RANS and high-fidelity data in the turbulence model[19-21].

There are still limitations of the current state-of-the-art research. Among these limitations, model extrapolation to unseen flow is the biggest challenges facing data-driven turbulence modeling, especially for models with high dimensional inputs that involve many flow features. It will be ideal to establish data-driven approaches that are generalizable to different or unseen scenarios, however, almost all previous work only demonstrate the predictive ability to limited flow cases that are similar to the training flow rather than validating the performance broadly for a wide range of flows. Even some studies[22] presented that the proposed model shows good agreement when generalized to both interpolated and extrapolated cases, mostly the reason is that there is no qualitative difference between training and predicting cases. Many researchers reported that data-driven modeling has satisfying performance on similar flows, but the performance decreases dramatically on flows that are significantly different from the training cases. For example, Ling and Templeton[12] reported that flow wake regions with very high turbulent intensity are labeled as extrapolation cases and have elevated error rates and poor prediction. In another study[13], the proposed tensor based neural networks (TBNN) is only shown to work adequately for similar flows, but not for the test cases which deviate significantly from the training flows in both domain geometry and Reynolds number. Geneva and Zabaras[23] built a library of five different flow cases, including converge-diverge channel, square cylinder, periodic hills, square duct, and tandem cylinders, in the hope to capture different flow physics. The used invariant neural network still proved difficult to train and yield satisfactory predictions for unseen flows and geometries. Wu[24] attempted to extrapolate the flow in a different geometry from a square duct to a rectangular duct and encountered less success compared to testing on the same geometry. In a similar study, Wang[19] also found that the improvement of the corrected Reynolds stresses for test cases with different geometries are not as drastic as in the scenario with the same geometry. Wu and his colleague[20] presented a comprehensive framework for augmenting turbulence models using physics-informed ML. To ensure the extrapolative capabilities of the learned function, the input variables are physically justified and properly normalized. They examined the model extrapolation capability by testing on 1) same geometry but different flow Reynolds number and 2) similar Reynolds number but different geometrical configurations. Their results show that a satisfactory predictive capability at lower Reynolds number does not necessarily guarantee a similar performance at a high Reynolds number. They concluded that existing data-driven turbulence models, including the one presented in their work, are still in their infancy and have shown only limited predictive capabilities.

These previous efforts suggest that data-driven turbulence modeling can interpolate in between extremes but is unable to extrapolate much beyond its training cases. It is more likely that ML can merely fit and interpolate rather than reveal much flow physics from existing data. It has been argued[25] that to ensure the predictability, ML should be used to learn the parameters or parametric functions within a traditional parameterization framework. ML algorithms with less interpretation, like the neural network and random forest, can be replaced with algorithms that are more interpretable like the symbolic regression and gene expression programming[15,26]. Because the known physics are hard coded, this could lead to better generalization capabilities and a reduction of the required data amount, but it cannot eliminate the burden of heuristically finding the framework equations. One can also argue that it has been proved for many decades that flow-specific tuning is inevitable for turbulence modeling and ML can be considered as an automatic tuning tool to replace the laborious modeling procedure. Since the extrapolation capability depends on the diversity of the training data, it seems fairly to claim that we can always use more comprehensive databases with various flow physics to expand the performance range. However, resources are limited to generate high quality databases. Therefore, it is of great value to answer these questions: What are good metrics that can quantitatively measure the dataset variability? To what extreme can data-driven turbulence modeling extrapolate and how to increase the predictability?

In this study we review previous efforts on RANS turbulence modeling and the model extrapolation, with a main focus on the popular choices of input features and ML models. Several potential metrics to measure the dissimilarity between training flows and testing flows are examined. Different ML models are compared to understand how the capacity or complexity of the model affects its behavior when there is training/testing dataset dissimilarity. Data preprocessing procedures which are robust to covariate shift, like transformation, resampling, and importance re-weighted likelihood, are studied to figure out whether it is possible to find projections of the data that attenuate the differences in the training and test distributions while preserving predictability. These findings are useful for future ML based turbulence models to evaluate their model predictability and provide guidance to systematically generate high-fidelity simulation database.

The paper is organized as follows. Section 2 reviews the ML assisted turbulence modeling, including ML modeling and input feature selection. Section 3 introduces the "flow over the bump" datasets that we use in this study. Section 4 summarizes the proposed dissimilarity metrics that determine how much the test flow is different from the training flows. Section 5 discusses data preprocessing procedures that can attenuate the dataset dissimilarity to improve model predictability. Section 6 concludes the paper.

## 2. ML assisted RANS turbulence modeling

### 2.1. RANS turbulence modeling

The modeling of Reynolds stress is the fundamental closure problem that is introduced when the Navier-Stokes equations are averaged with respect to time. The RANS momentum equation is written as:

$$\langle u_j\rangle \frac{\partial \langle u_i\rangle}{\partial x_j} = \frac{\partial}{\partial x_j}\left[-\frac{\langle p\rangle}{\rho}\delta_{ij} + \nu\left(\frac{\partial \langle u_i\rangle}{\partial x_j} + \frac{\partial \langle u_j\rangle}{\partial x_i}\right) - \langle u_i' u_j'\rangle\right] + \langle g_i\rangle \tag{1}$$

How to model the unclosed Reynolds stress anitropic tensor $\langle u_i' u_j' \rangle$ using the available flow field is the most important issue in RANS. Many RANS models rely on the Boussinesq hypothesis, as it yields accurate results in many simple shear flows and the additional eddy viscosity conveniently aids numerical convergence. The Boussinesq hypothesis is:

$$\langle u_i' u_j' \rangle = \frac{2}{3}\delta_{ij}\kappa - \nu_t \left( \frac{\partial \langle u_i \rangle}{\partial x_j} + \frac{\partial \langle u_j \rangle}{\partial x_i} \right) \tag{2}$$

$$s_{ij} = \frac{1}{2}\frac{\kappa}{\varepsilon}\left( \frac{\partial \langle u_i \rangle}{\partial x_j} + \frac{\partial \langle u_j \rangle}{\partial x_i} \right) \quad \omega_{ij} = \frac{1}{2}\frac{\kappa}{\varepsilon}\left( \frac{\partial \langle u_i \rangle}{\partial x_j} - \frac{\partial \langle u_j \rangle}{\partial x_i} \right) \tag{3}$$

The turbulent kinetic energy is defined as half the trace of the Reynolds stress tensor, written as:

$$\kappa = \frac{1}{2}\left( \langle u_i'^2 \rangle + \langle u_j'^2 \rangle + \langle u_k'^2 \rangle \right) \tag{4}$$

The majority of the popular RANS turbulence models use the Boussinesq approximation and introduce one or two additional transport equations. However, it relies on several underlying assumptions that are violated in many common flows. The turbulence related transport equations contain a set of tunable constants which are calibrated using a small number of simple test cases such as homogeneous turbulence and thin-shear flows. In addition, the models include pre-specified functional forms for the closure terms that are typically chosen using the knowledge of the flow physics and the intuition of the turbulence model developers. Given this development process, it is not surprising that accuracy diminishes as the model is applied to problems which deviate from the set of calibration cases. Some recent encouraging efforts have proposed more sophisticated models[27], however, a paradigm shift is needed to effectively eliminate these limitations.

## 2.2. ML models

Figure 1 illustrates the work flow of the ML-assisted RANS simulation. To prepare the training dataset, the flow features are calculated from RANS simulations, whereas the Reynolds stresses are calculated from high-fidelity DNS/LES simulations. ML models are trained to establish a functional relation between the input (flow features) and output variables (Reynolds stresses or related parameters). The predicted Reynolds stresses are then substituted into the RANS solver to obtain corrected flow field.

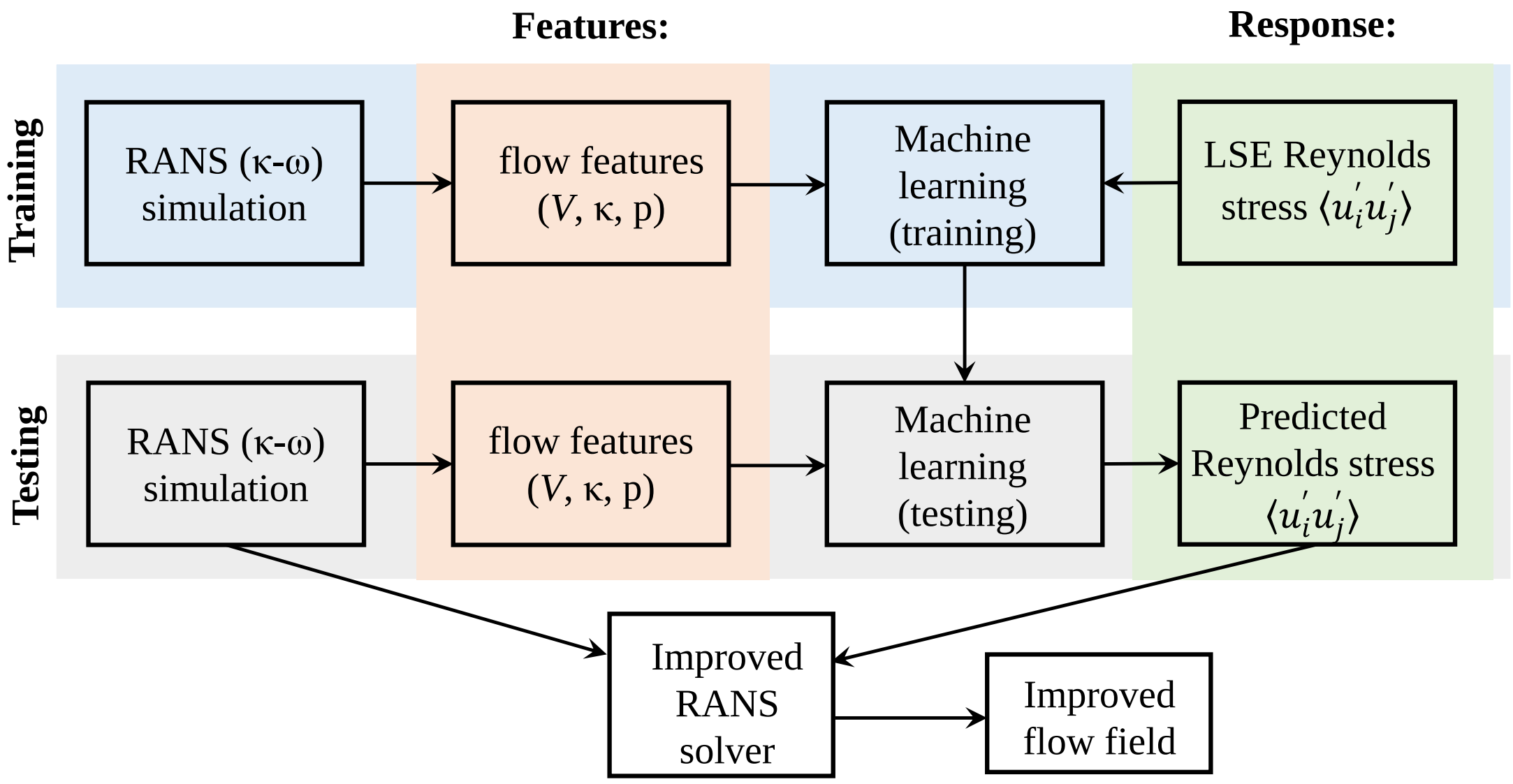

*Figure 1. Schematically illustrate the workflow of ML-assisted RANS simulation.*

There are a variety of ML-based approaches for turbulence modeling. For example, a feed forward neural network is employed in establishing a functional relation between the mean flow quantities and the Reynolds stress tensor.[28,29] Besides of the simple feed-forward neural network, a tensor based neural network (TBNN) was reported to have significant performance increase.[13,30] The theoretical foundation of the tensor based neural network is the non-linear eddy viscosity model developed by S. Pope.[31] In this model, the Reynolds stress anisotropic tensor is expressed as a function of the normalized mean rate-of-strain tensor and rotation tensor. To enforce invariance to the coordinate transformation, the neural network is used to learn the tensor basis coefficients which are functions of five invariants. The basis of invariants of mean strain rate and rotation rate tensors with respect to the orthogonal group can be constructed using tables reported in paper[14] which enumerate the relevant invariants for a symmetric and anti-symmetric tensor. The random forests is identified as the optimal approach to model RANS discrepancies based on the criteria of high dimensionality and interpretability.[19] In our work, we examine four supervised learning algorithms: feedforward neural network (NN), random forests (RF), support vector machine (SVM), and gradient boosting (GB), in the hope to find the optimal approach to build the functional form that is most suitable for turbulence modeling.

With the ML algorithms identified, another challenge is to identify a set of mean flow features based on which ML algorithms can be constructed, the so-call feature selection. Feature selection is the process by which relevant input features for an error metric are chosen. In principle there are a huge number of possibilities and combination of input variable. Wang et al. employed a set of 10 invariant features for building the random forest regressor for Reynolds stress discrepancies.[19] Duraisamy used a set of parameters including the full velocity gradient tensor, the transported scalars in transport equations, and all non-dimensional parameters that appear in the RANS model.[17] Ling et al. proposed a systematic methodology for constructing an input feature set. Specifically, given a finite collection of raw inputs (i.e., tensors or vectors), a finite integrity basis of invariants can be constructed, and any scalar invariant function of the raw inputs can be formulated as a function of the corresponding invariant basis.[14]

While the selection of these input features heavily relied on physical intuition and reasoning, it is inevitable that the selection introduces human judgement. Indeed, any combination of features could be used. However, it is possible that the input set may exclude important physical information if some key invariants are omitted. Therefore, it makes more senses to identify the raw input vectors and tensors which are related to the mean flow at the first step. In this work, the selected input variables are three flow features: velocity, pressure, and turbulence kinetic energy (TKE). These raw mean flow features are assumed to represent the important physical characteristics of the mean flow, which are also used as important elements for traditional turbulence modeling. It should be noted that the current work focuses on the assessment of similarity between the training and the test flows with the proposed measurement metrics in this work is applicable to other choice of flow feature spaces.

## 3. Flow cases for training and testing

The turbulence model was evaluated for a family of five 2-D bumps of various heights (the "flow over the bump" dataset mentioned in the Introduction). The large eddy simulation (LES) dataset is a well-known NASA validation case for which the settings and computational grids can be found in previous literature.[32] The five bump cases have heights $h$ = 20 mm, 26 mm, 31 mm, 38 mm, and 42 mm, with the same Reynolds number of 2500. With these five cases, 10 pairs of training and testing dataset pairs can be obtained. Table 1 shows the 10 scenarios tested in this paper. Here, bump 20 stands for the case with bump height h = 20mm. Figure 2 shows the contour plots of velocity x, velocity y, pressure, and TKE with bump height $h$ = 20 mm. This case study is perfect to demonstrate the prediction performance for which the training flow and the test flow have different geometrical configurations. This scenario is also very realistic in the context of using RANS simulation to support engineering design and analysis that, the training data are more likely to be available for a few flows with specific Reynolds numbers and geometries, but predictions are needed for the similar flows yet with modified geometries. For example, studies are conducted to explore the feasibility of machine learning to assist airfoil design by interpolating unstructured shape parameters.[22,33] All airfoils have similar shapes with slightly different geometry configurations including attach angle, leading edge and trailing edge, and thickness to chord ratio, etc. To predict testing cases with totally different geometries from the training cases is very challenging and less successful.

*Table 1. Ten combinations of flow cases tested.*

| | Case1 | Case2 | Case3 | Case4 | Case5 | Case6 | Case7 | Case8 | Case9 | Case10 |
|---|---|---|---|---|---|---|---|---|---|---|
| **Train** | Bump 20 | Bump 20 | Bump 20 | Bump 20 | Bump 26 | Bump 26 | Bump 26 | Bump 31 | Bump 31 | Bump 38 |
| **Test** | Bump 26 | Bump 31 | Bump 38 | Bump 42 | Bump 31 | Bump 38 | Bump 42 | Bump 38 | Bump 42 | Bump 42 |

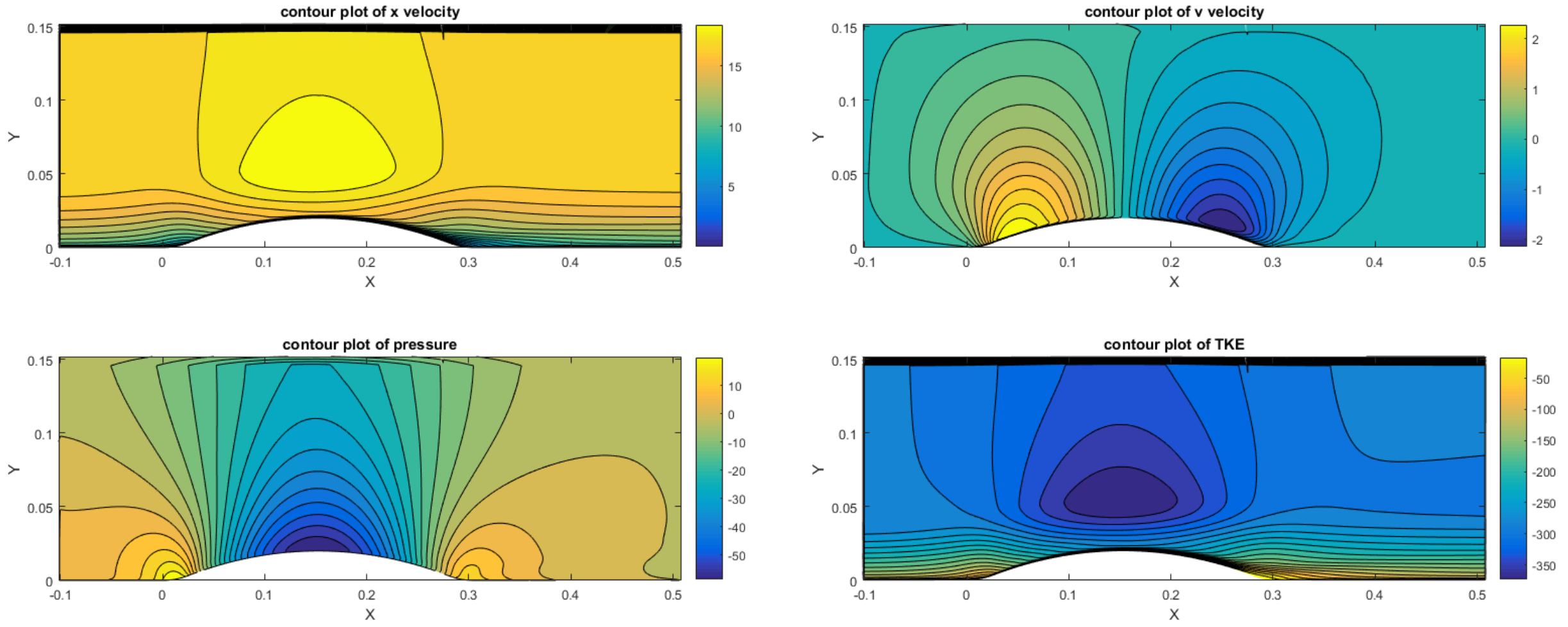

*Figure 2. Contour plots of velocity x, velocity y, pressure, and TKE with bump height h = 20 mm.*

For each bump dataset, the grid points are 591 × 160 in x and y direction, so both the training and testing have 94560 samples. The hyperparameter tuning library, *Tune*[34], is used to find optimal number of hidden layers, number of nodes, and dropout rate. The network training is stopped when the model error of the hold-out validation dataset (20% training) stopped decreasing. The network is found to have best prediction when the nodes and layers are [4, 32, 32 8, 8, 8, 3], and the dropout out is 0.2. Figure 3 shows the predicted Reynolds stresses versus real values for three different pair of datasets when using NN as the modeler. There are three components in the Reynolds stress for 2D simulation. The first two components in Reynolds stress tensor are easier to predict due to the largest variation. On the other hand, the third component is difficult to be approximated as the variation is relatively small. We gradually increase the training difficulty deliberately to test the model performance under various dataset dissimilarity. The result in Figure 3 suggests that better predictive performance is obtained when the training flow and the prediction flow are more similar. The first case has acceptable performance while the second and third cases fail to generate satisfactory prediction. This failure comes from the fact that the training and testing flow are significantly different that it beyond the predictability of the built model. In flow physics the lowest bump had no separation, while the highest bump produced a marginal separation. Bumps with h = 20, 26, 31 have trivia separation but bumps with h = 38 and 42 have considerable separation. We will introduce how to measure this dissimilarity from a statistical prospect in the next section.

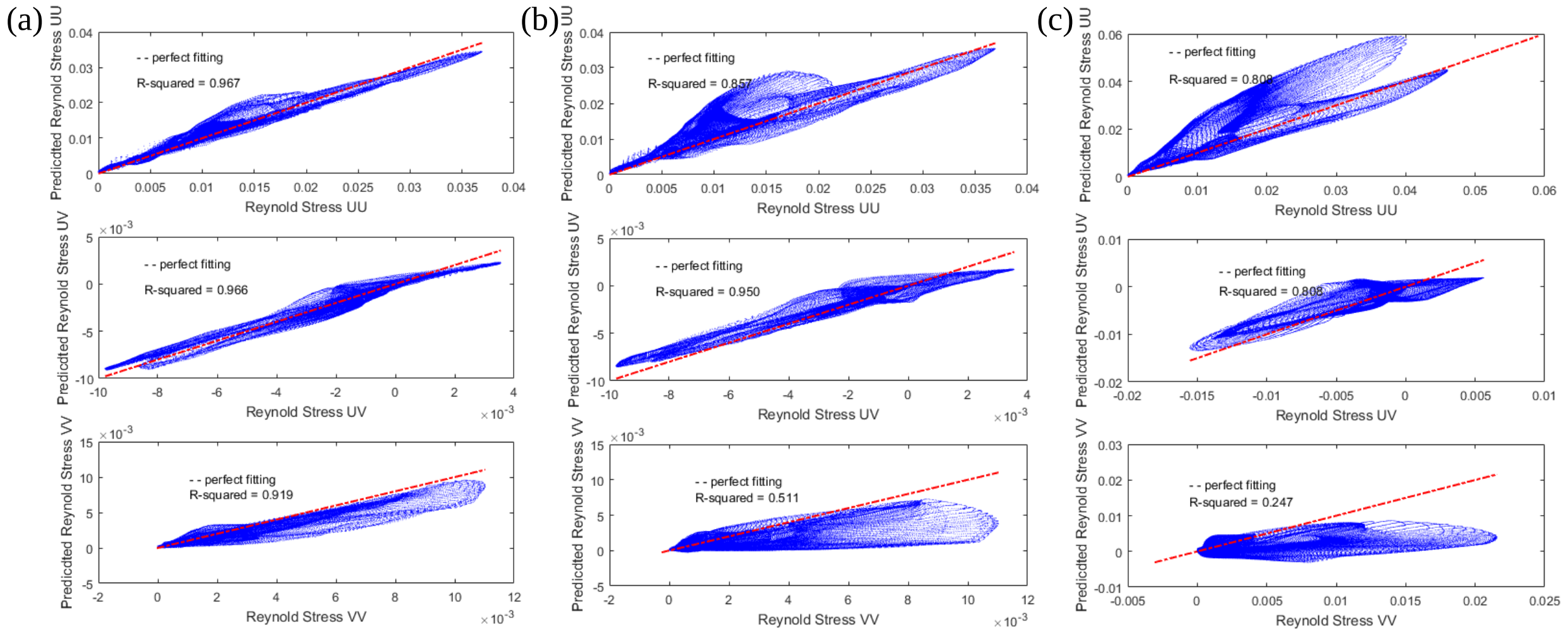

*Figure 3. Testing results for three combinations of training and testing dataset pairs: (a) case 1; (b) case 2; (c) case 4. The R-squared, which is the proportion of the target variance that is predicted from the features, is reported in each case. The red dashed line is the perfect fitting y = x as a guide to the eye.*

Figure 4 shows the root mean squared error (RMSE) and mean absolute error (MAE) for four different ML algorithms. The hyperparamter tuning of SVM, GB, and RF is accomplished using GridSearchCV.[35] From the figure, all four methods have very close performance, neural network has slightly better performance with lower MAE than gradient boosting and random forest in most cases, indicating that neural network has better generalization ability particularly in this high-dimensional nonlinear regression study.

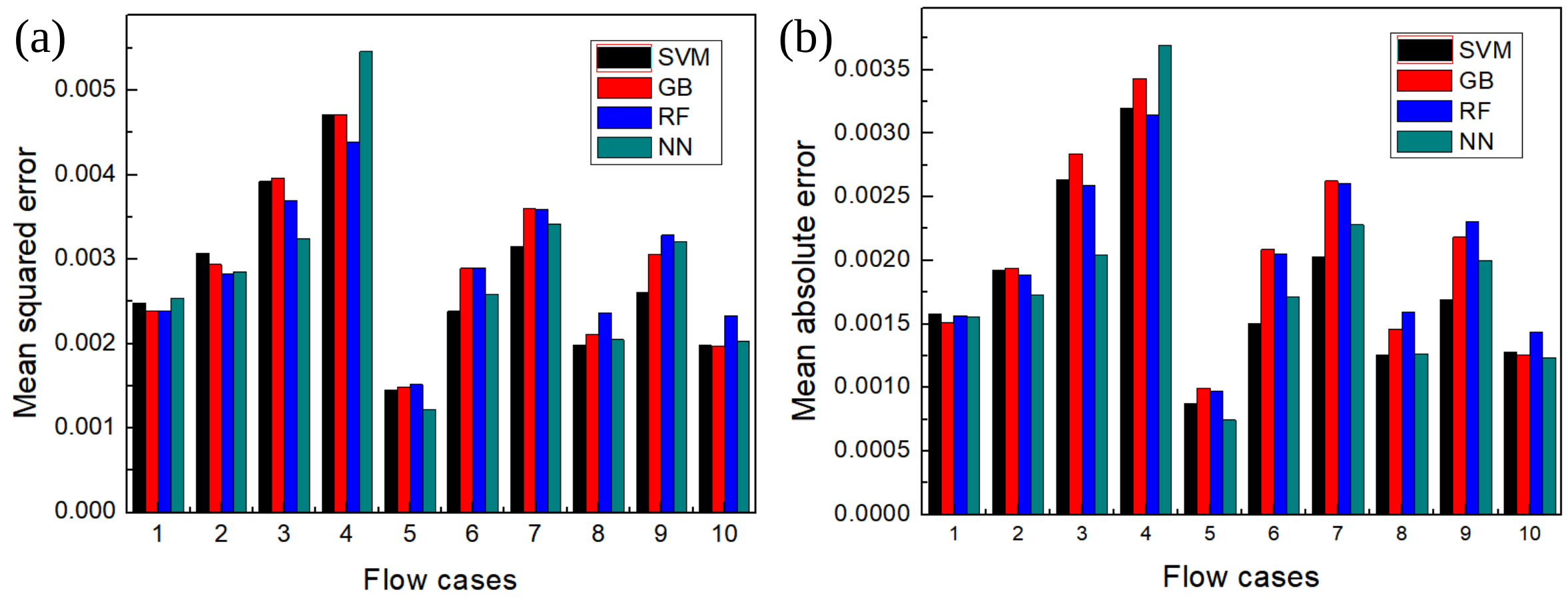

*Figure 4. Root mean squared error and mean absolute error for four different ML algorithms.*

## 4. Dissimilarity metrics

The unsatisfactory modeling extrapolation is due to the fact that unseen testing flow datasets usually have different distribution than the training dataset. Most predictive models are constructed under the assumption that the training/testing datasets are governed by the exact same distribution. For example, flow dataset of attached boundary layer has very different distributions than flow dataset with separation. Also, if none of the training cases have shock waves, it would not be surprising if the model performed poorly in a test set with a shock wave. While fluid scientists can distinguish flows with different physics and evaluate the relevance between training/testing cases using domain knowledge, it is still valuable to measure readily available high-fidelity CFD databases from a statistician's view. Some metrics are examined to quantify the dissimilarity of two datasets considering both location distance and correlation structure difference.

### 4.1. Dissimilarity decomposition

The quantification of the dissimilarity is critical as it can set reasonable expectations of model prediction accuracy. The dissimilarity of two dataset $\bar{X}$ and $\bar{Y}$ can be decomposed into three components that represent the differences in location, rotation, and "shape", respectively.[36] The location component is defined as the Euclidean distance, which measures the ordinary straight-line distance between the centroids of each dataset, or the Mahalanobis distance, in which the distance is scaled by component covariance. The rotation component is the measure of the angles between their principal components, it is the degree that one dataset must be rotated to align with the other principal component. The "shape" component accounts for the difference in the shape of the dataset distribution. Two variables in a dataset can be spread around its mean value or aligned in the same direction. Figure 5 schematically illustrates the dissimilarity metrics for four synthetic datasets. (a) similar datasets drawn from the same distribution; (b) datasets with location distance; (c) datasets with rotation distance; (d) datasets with shape distance.

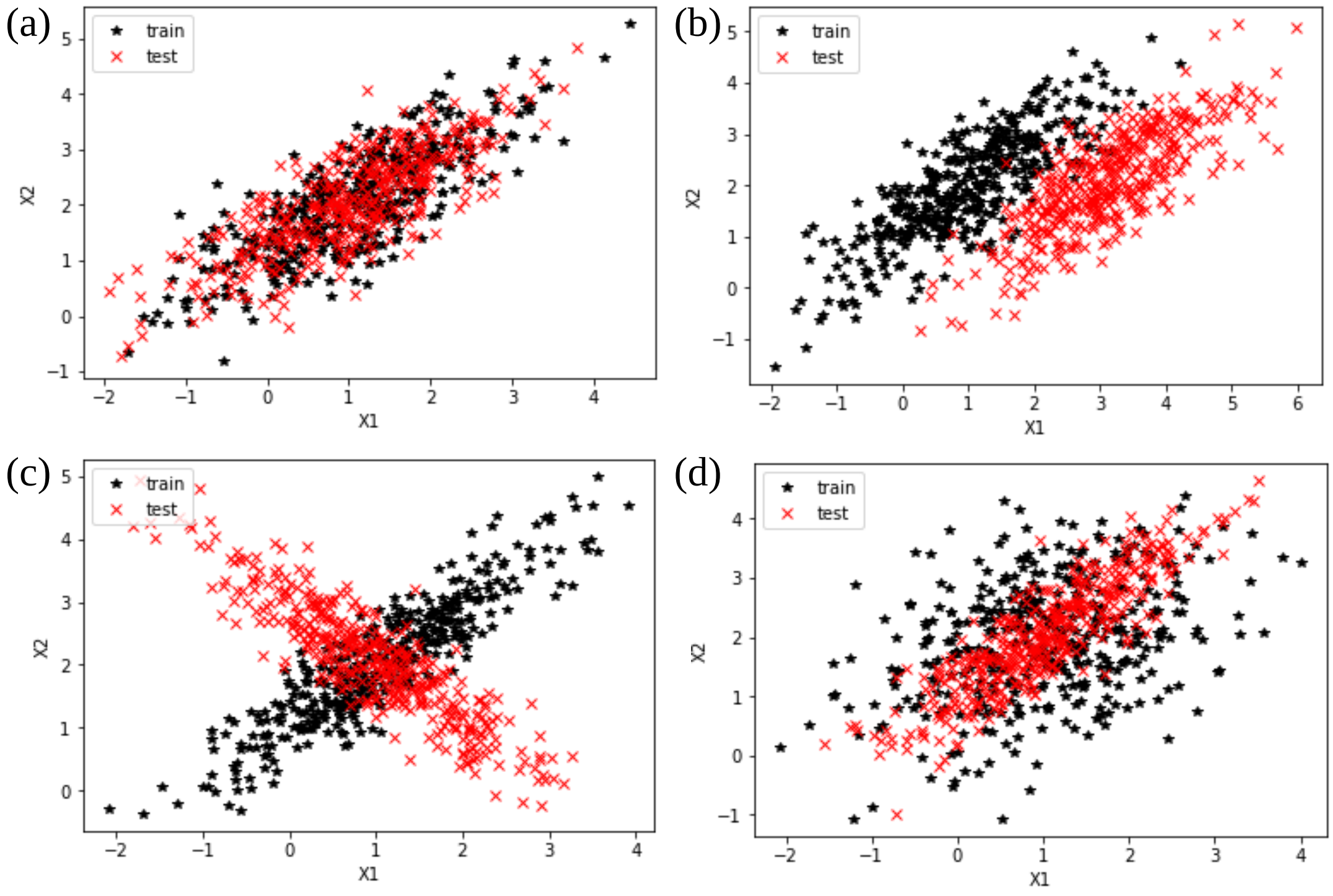

*Figure 5. The illustration of dissimilarity metrics for four synthetic datasets. (a) similar datasets drawn from the same distribution; (b) datasets with location distance; (c) datasets with rotation distance; (d) datasets with shape distance.*

**Location distance**: This metric was presented to measure the extrapolation in different flow dataset by quantifying the centroid distance between the training and testing datasets.[12,37] Euclidean distance is commonly used to measure the difference between two points or a point and a distribution. However, Euclidean distance will not work when features in high dimensional datasets are correlated to each other or not equally weighted. A statistical metric called the Mahalanobis distance measures not only the distance from the center of mass, but also the direction. The distance is scaled by the covariance matrix to account for the fact that the variances in each principle direction are different. Mahalanobis distance is defined as $D_d^2(X,Y) = (\mu_X - \mu_Y)C^{-1}(\mu_X - \mu_Y)^T$, where $\mu$ is the mean and $\boldsymbol{C}$ is the covariance matrix. If the variables in dataset are strongly correlated and the covariance is high, dividing the distance by a large covariance will effectively reduce the distance. The major drawback is that it requires the inversion of the covariance matrix, which can be computationally restrictive if variables are highly correlated.

**Rotation distance**: The rotation distance measures the degree to which the testing dataset must be rotated so that its principle components point in the same direction as the training dataset. The principal components of a dataset are the set of orthogonal vectors such that the first vector points in the direction of greatest variance in the data, the second points in the orthogonal direction of the second greatest variance in the data, and so on. The two datasets are similar to each other if their principal components are aligned with each other. The rotation distance measures the summation of the angles between principal components as $D_r(X,Y) = trace\left(cos^{-1}\left(abs(V_X V_Y^T)\right)\right)$, where $V$ is the principal components vector.

**Shape distance**: Datasets with different distribution present different "shapes". The kernel density estimation (KDE) distance is commonly used to measure how on probability distribution is different from a second reference probability distribution. The probability distributions usually are not known a priori, which means they need to be estimated from data. KDE is a non-parametric way to approximate the probability density function of multivariate variable without any pre-assumption of the distribution, it uses a mixture consisting of one Gaussian component *per point*, resulting in an essentially non-parametric estimator of density. The KDE is defined as $\Phi(\mathrm{x}) = \frac{1}{nh}\sum_{i=1}^{n} K\left(\frac{x-x_i}{h}\right)$, where $K$ is the kernel function and $h$ is the bandwidth. The free parameters of kernel density estimation are the kernel, which specifies the shape of the distribution placed at each point, and the kernel bandwidth, which controls the size of the kernel at each point. The choice of bandwidth within KDE is extremely important to find a suitable density estimate: a too narrow bandwidth leads to a high-variance estimate (i.e., over-fitting), where the presence or absence of a single point makes a large difference. A too wide bandwidth leads to a high-bias estimate (i.e., under-fitting) where the structure in the data is washed out by the wide kernel. A Gaussian kernel is commonly used, and the bandwidth is determined empirically via a cross-validation approach. The procedure of obtaining the KDE distance between two datasets with different lengths is first to obtain the kernel density estimated distribution $\boldsymbol{\Phi_1}$ of the training dataset and the probability $\boldsymbol{P_1}$ of all training data points; then to obtain the kernel density estimated distribution $\boldsymbol{\Phi_2}$ of the testing dataset and get the probability $\boldsymbol{P_2}$ evaluation on the training dataset; the KDE distance is then calculated.

Figure 6 summarizes the three dissimilarity measures in different pairs of flow cases. Here, these distances are not normalized, which means the absolute value does not have a meaning. It is the comparison between different cases are the information we look for. The left figure shows that all three metrics are aligned when describing the dissimilarity. It is intuitive to see that Case 4 (bump h = 20 mm for training and h = 42 mm for testing, these two flows are very different) has the largest dissimilarity while case 5 (bump h = 26 mm for training and h = 31 mm for testing, there two flows are similar) has the smallest dissimilarity.

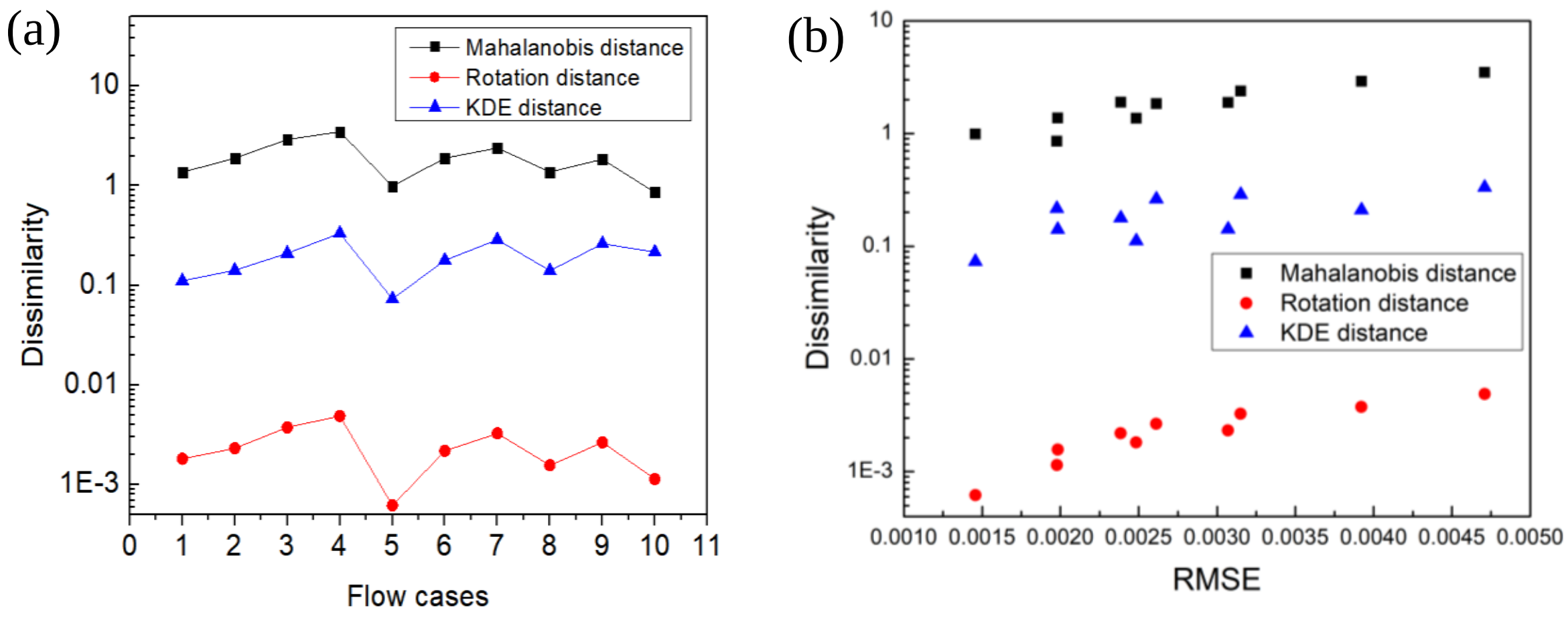

*Figure 6. (a) Three dissimilarity measure in different pairs of flow cases; (b) The relationship between the mean error of the prediction of Reynolds stress anisotropy and dissimilarity metrics.*

## 4.2. Dissimilarity vs. predictability

We use different pairs of training/test dataset in table 1 to predict the flow over bump. By analyzing the mean prediction error of Reynolds stress anisotropy based on different training sets, we can see that there is a positive correlation between mean prediction error of Reynolds stress anisotropy and the dissimilarity metrics in Figure 6(b). This result suggests that better predictive performance is obtained when the training flow and the prediction flow are more similar. For example, the predictive performance is the best when both the training case and the testing case are from the same LES dataset when $h$ = 31. In contrast, the worst performance occurs when training case $h$ = 42 while testing on $h$ = 31. By visualization it is intuitive to see that the two cases $h$ = 42 and $h$ = 31 are very different, the downstream separation in $h$ = 42 are much more severe that when $h$ = 31. The lowest bump had no separation, while, the highest bump produced a marginal separation.

## 5. Dissimilarity attenuation

While poor model performance is usually expected for a ML model when conducting extrapolation, the extrapolation capability is highly desirable and even critical for turbulence modeling, as dataset dissimilarity is common in turbulence modeling. It is of extreme importance to find projections of the data that can attenuate the differences while preserving predictability. To build models robust enough to account for distributional shift is not easy and still on-going research, however, there are data preprocessing schemes that can help improve model extrapolation, like transformation, resampling, and importance re-weighted likelihood.[38] Many previous efforts have been tried to tackle this so called domain adaption problem.[39,40,40,41,41,42,42] This section tries to find ways of improving the performance and also to determine the situations where a particular method results in a superior performance.

### 5.1. Data transformation

Data transformation, like nondimensionalization, standardization, threshold, and transformation may help to improve the extrapolation performance. For example, researchers noticed that dimensional quantities are not an appropriate choice for the input feature vector to the ML algorithm.[12,18,20] Wu and his colleagues stated that the variables must be normalized properly to ensure extrapolative capabilities of the learned function.[20] It is necessary to un-dimensionalize the features by relevant local quantities that are representative of the state of turbulence. Besides of dimensionless treatment, the flow data tended to vary strongly even in order of magnitude.[13,23] To tackle this, the data are scaled linearly into a predefined range to reduce the symptoms, while the distribution characteristic of the data is preserved. The most challenging part is that flow datasets usually are highly skewed and non-normally distributed with multiple dimensions. While ML algorithms like neural network estimator are model free and require no assumption on data distribution, many studies reported a significant performance enhancement when we modify the distribution characteristics of the dataset, so input features are skewed.[43,44] Power transformation can reduce the skewness, whereas the Box-Cox procedure is commonly used to determine an appropriate power transformation. The standardization and power transformation are used in the flow data preprocessing in this study to attenuate the dataset dissimilarity.

### 5.2. Resampling

While data transformation can attenuate the dissimilarity by reducing the location distance, the distribution difference of the data is still preserved. Resampling methods can reweight instances in the training data so that the distribution of training instances is more closely aligned with the distribution of instances in the prediction dataset. This is accomplished by providing more weighting to an instance in the training dataset that are similar to the prediction dataset. The optimal choice of the weight function is asymptotically shown to be the test-to-training density ratio.[38] The densities ratio characterizes how much more likely an instance is to occur in the test sample than it is to occur in the training sample. The importance of the weighting scheme is intuitively understandable. If the probability of seeing a particular training instance in the prediction is very small, then this instance should carry little weight during the training process. In practice, the approach to resampling would be to first estimate the training and test densities separately and then estimate the ratio of the estimated densities of test and train. The estimated densities act as resampling weights for each instance in the training data. We adopted the Kullback-Leibler importance estimation procedure (KLIEP) to obtain the resampling weights to attenuate the training/testing dissimilarity, the detailed implementation of KLIEP can be found in previous paper.[45]

We re-measured the three dissimilarity metrics after the data transformation and data resampling. From Figure 7 it seems that the Mahalanobis distance and rotation distance increases after the preprocessing, but the KDE distance reduces drastically after resampling, rendering a distribution matching between training/testing dataset. A drawback of this preprocessing is that the rotation distance increases after resampling, it is our hope that this rotation metric is not dominating and will not affect the training performance.

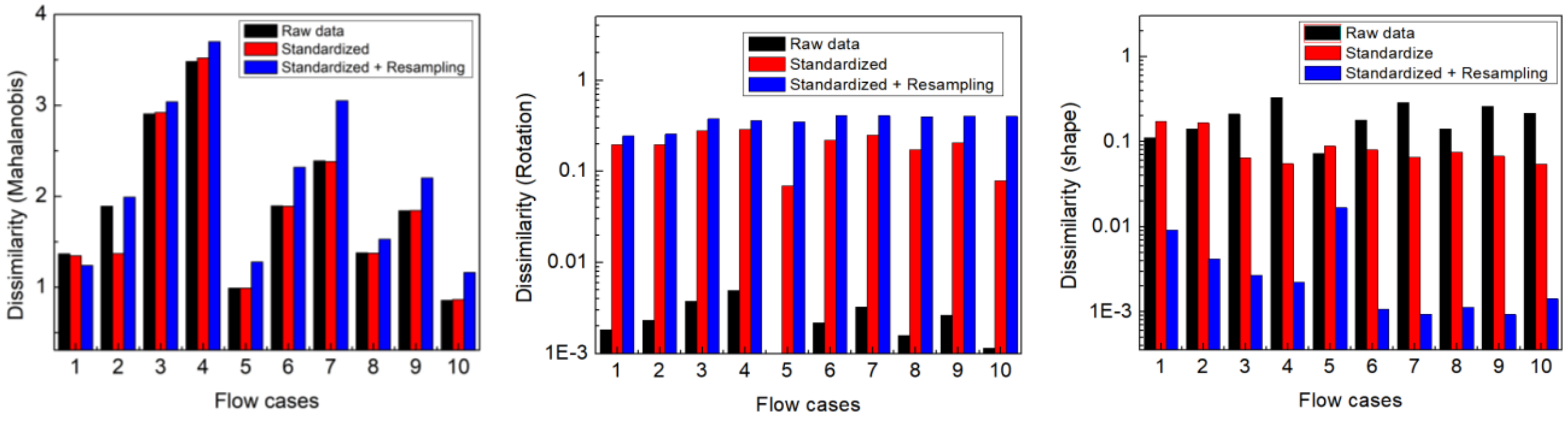

*Figure 7. Comparison of the three dissimilarity metrics among raw data, standardization, and resampling. It shows that the (standardization+resampling) can drastically reduce the KDE distance, rendering a distribution matching between training/testing dataset.*

Another reason for resampling is that the flow datasets are very unbalanced that regions with rich physics are usually severely under-represented. These datasets include a large bulk region with very similar flow structures, making many training points redundant. In flow over flat plate, for example, points that have the largest predictive variance are these points close to wall in turbulence boundary layer, while the majority points in the outer layer contain less information. It is found in our flow cases that if we use all the flow dataset for training, ML models will over-smooth the entire prediction field in an attempt to cater towards the large number of data points in the bulk region, while the underrepresented rich physics is mispredicted. The errors are largest in regions where the adverse pressure gradient and separation occurs in the downstream. Although the entire flow domain seems to contain more relevant information, in practice it is more desirable to resample sub-datasets that contains more information in order to get improved model performance.

Upper-sampling of under-represented samples that contains most information can potentially increase the predictability.[46-48] In Geneva's paper[30], only points that have the largest predictive variance for the Reynolds stress anisotropic component are selected for training. The resampling is found helpful to prevent the model from focusing too much on the bulk flow at which the prediction quality is poorer. Chawla[48] proposed an interpolation strategy, the Synthetic Minority Over-Sampling Technique (SMOTE), to create synthetic examples that have rare but important target values. A resampling approach is proposed to change the distribution of the given data set to attenuate the imbalance between the rare target cases and the most frequent ones. The strategy is to randomly select one of its *k*-nearest neighbors from the set of observations, then a new example is created with interpolated values of the two original cases. To decide which points are "rare", the standardization and KLEIP resampling are conducted ahead this SMOTE technique so the datasets are well centered around zeros. The data points with absolute value > 2 are considered as "rare" cases, while points with absolute value < 2 are treated as "normal" cases. This classification is decided by the user and can be adjusted accordingly to obtain optimal modeling performance. The SMOTE technique is open-sourced as a python toolbox[49]. The results comparison before and after pre-processing are presented in Figure 8. The errors are consistently reduced for all cases after the sampling schemes are applied. These extensive set of experiments provide empirical evidence for the superiority of proposed schemes for these particular regression tasks.

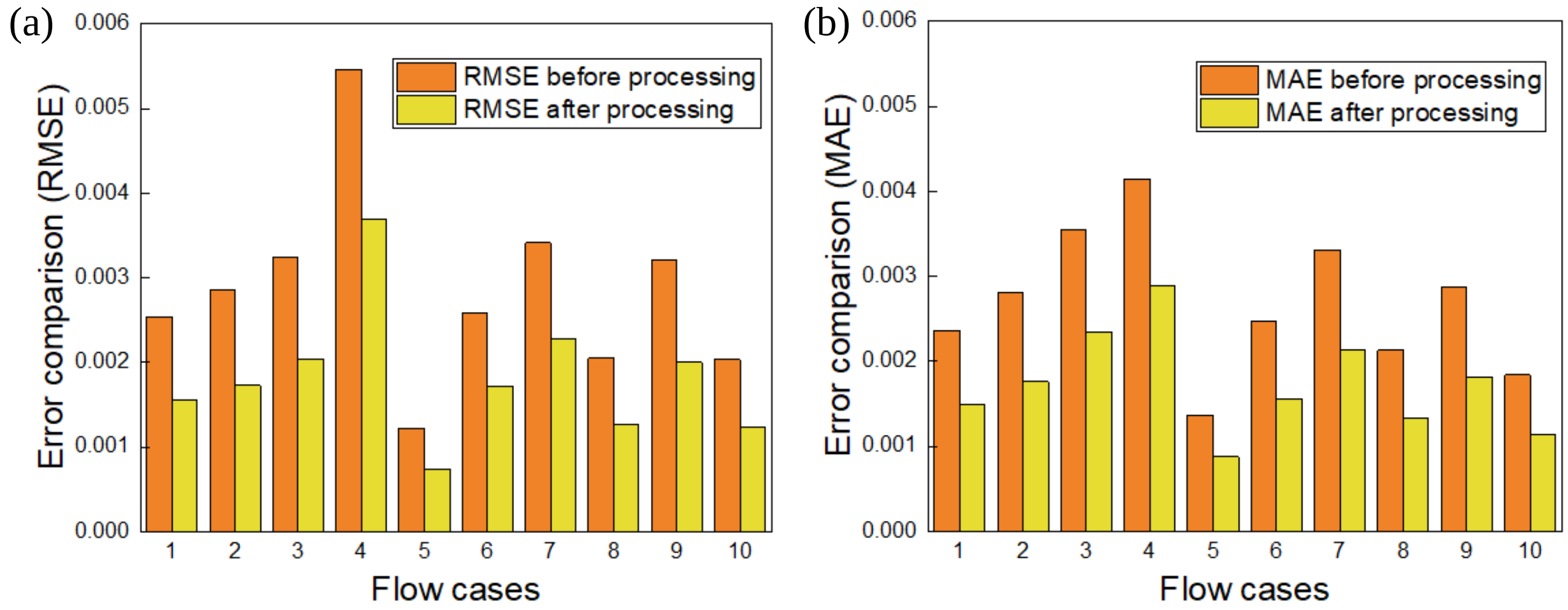

Figure 8. Errors comparison before and after the pre-processing. (a) RMSE, (b) MAE.

## 6. Conclusion

In this study we reviewed previous efforts on RANS turbulence modeling and the model extrapolation, with main focus on examining several methods to measure and attenuate the dissimilarity between training and testing flows. Data preprocessing schemes which are robust to covariate shift, like normalization, transformation, and importance re-weighted likelihood, are studied to find dataset projections that can attenuate the differences in the training and test distributions while preserving predictability. This dissimilarity measurement study can work as guidance on the choices of preprocessing schemes for ML assisted turbulence modeling to achieve better prediction performance. These findings are also useful for future ML based turbulence models to evaluate their model predictability and provide guidance to systematically generate high-fidelity simulation database. The dissimilarity metrics ideally will provide a direct indication of the need for additional data and drive the direction for further data-collection activities. For example, Wu stated that at high Reynolds number even Reynolds stresses with average errors below 0.5 % will lead to the propagated mean velocities with large errors (up to 35 %).[50] If one wants to keep the velocities error at a lower level, more training datasets are required to provide sufficient dataset diversity to restrain the dissimilarity. To attenuate the dissimilarity, a distribution matching framework is used to align the distributions of the source domains and target domain. These modifications also allow the regression tasks to have better accuracy in forecasting under-represented extreme values of the target variable.

Future efforts can focus on finding more suitable assessment metrics. In all proposed metrics, only the inputs of the training and test data are used, and the information of the response is not needed. While these three metrics to some extend can measure the covariate data shift, it does not measure the prior probability shift and concept shift. These studies lack in considering and explaining the effect of the distribution of the target dataset which is as much decisive as input data on neural learning. Another future study is that embedding these error metrics with physical constrains into the loss function can potentially increase the generalization ability. Incorporation of known physical constraints on the data-driven models will greatly reduce the demands on data and promote generalization.

**ACKNOWLEDGMENTS**

This work utilizes resources supported by the National Science Foundation's Major Research Instrumentation program, grant #1725729, as well as the University of Illinois at Urbana-Champaign.